\documentclass[twocolumn,preprintnumbers,amsmath,amssymb]{revtex4}
\usepackage{graphicx}% Include figure files
\usepackage{dcolumn}% Align table columns on decimal point
\usepackage{bm}% bold math

\begin{document}

\title{High resolution atomic coherent control via spectral phase manipulation of an optical frequency comb}

\author{Matthew C. Stowe, Flavio C. Cruz,* and Jun Ye}
 \affiliation{JILA, National Institute of Standards and Technology and University of Colorado\\ and Department of Physics, University of Colorado, Boulder, CO 80309-0440}
% \affiliation{$^{2}$Instituto de Fisica Gleb Wataghin, UNICAMP, Campinas, SP, 13083-970, Brazil}
\date{\today}

\begin{abstract}
We demonstrate high resolution coherent control of cold atomic
Rubidium utilizing spectral phase manipulation of a femtosecond
optical frequency comb. Transient coherent accumulation is directly
manifested by the enhancement of signal amplitude and spectral
resolution via the pulse number. The combination of frequency comb
technology and spectral phase manipulation enables coherent control
techniques to enter a new regime with natural linewidth resolutions.

\end{abstract}

\maketitle

The introduction of phase-stabilized optical frequency combs has
revolutionized the field of precision atomic and molecular
spectroscopy \cite{Udem1,Ye1}. In our previous work we used direct
frequency comb spectroscopy (DFCS) to demonstrate high resolution
measurements of one- and two-photon transitions in cold $^{87}$Rb
\cite{Marian1,Marian2}. This new technique has the significant
advantage that the comb frequencies may be absolutely referenced,
for example to a Cesium atomic clock, enabling precision
spectroscopy over a bandwidth of several tens of nanometers. The
time-domain phase coherence offered by the optical frequency comb
has stimulated research in areas of atomic clocks \cite{Diddams1,
Ye2}, optical and radio frequency transfer \cite{Holman1}, and more
recently high harmonic generation \cite{JJones1,Hansch2}. In
parallel, the field of coherent control of atomic and molecular
systems has seen advances incorporating high power femtosecond laser
sources and pulse shaping technology. This has allowed for
demonstrations of robust coherent population transfer via adiabatic
passage techniques \cite{Bergmann1,Noordam1}, coherent control of
two-photon absorption \cite{Silberberg1,Noordam2,Girard1},
resolution enhancement of coherent anti-Raman scattering
\cite{Silberberg2}, and progress towards cold atom photoassociation
\cite{Matthais}. It is under this exciting context that we combine
the femtosecond comb and spectral phase manipulation with the aim
towards coherent control at the highest possible spectroscopic
resolution.

In this Letter, properties of the transient coherent accumulation
effect, including spectral phase manipulation, are explicitly
demonstrated via a finite number of phase-stabilized femtosecond
pulses. Specifically, we measure atomic transition linewidth and
population transfer versus the number of applied phase-coherent
femtosecond pulses and as a function of linear frequency chirp. The
basic idea of multi-pulse coherent accumulation is that the
amplitude for excitation of a specific atomic energy level may be
increased significantly with the number of femtosecond pulses, and
can be done such that other non-resonant states remain unexcited,
thus enabling high state selectivity. We show that tuning the comb
leads to destructive quantum interference of the two-photon
transition amplitude only observable in this multi-pulse context and
may be useful for eliminating nonlinear absorption. Positive and
negative linear frequency chirp is then applied to the comb modes
and thus the two-photon transition amplitudes. It is shown that the
role of chirp in the single pulse case is no longer necessarily
applicable in these multi-pulse experiments since the atomic
coherence persists between femtosecond pulses. The combination of
pulse shaping with the femtosecond comb is shown to increase the
signal of a two-photon transition at a specific chirp while
maintaining high resolution.

The spectrum of a train of femtosecond pulses is a set of
approximately $10^6$ comb modes equally spaced by the repetition
frequency ($f_r$) of the mode-locked laser. The frequency of each
comb mode is given by the relation $\nu_{\tiny{N}}= f_o+ N\times
f_r$, where $f_o$ is the carrier-envelope offset frequency and $\it
N$ is an integer mode number of the order $10^6$ \cite{Ye1}. The
frequencies of the comb modes are independent of the spectral phase
provided all pulses in the train have the same spectral phase. For
the two-photon transitions we investigate, there are of the order
2$\it N$ different transition amplitudes from the ground to the
excited state. It is instructive to consider the frequency spectrum
of the laser field consisting of a finite number of pulses. In this
work we use from one to approximately 100 pulses for atomic
excitation. The linewidth of the laser field after a single pulse is
approximately 30 nm, and after two pulses the spectrum is
sinusoidally modulated with a spacing of $f_r$ between peaks. In
general, a comb mode lineshape is described by a $sinc^2$ function
of the pulse number. As the linewidth narrows, the peak power of any
comb mode grows as the number of pulses squared. This amplitude and
spectral resolution enhancement presents a clear distinction from
previous experiments employing single pulse excitation.

To explore the transient coherent accumulation effects, we use a
sample of $^{87}$Rb atoms laser cooled and trapped in a
magneto-optic trap (MOT). This allows for approximately 50 pulses to
coherently interact with the atomic system via the $5S_{1/2}$ F=2 to
$5P_{3/2}$ F=3 to $5D_{5/2}$ F=4 two-photon transition (Fig. 1(a)).
We use a Kerr lens mode locked Ti:Sapphire laser which operates at a
center wavelength of 778 nm with a 30 nm wide spectrum and 100 MHz
repetition rate. This produces transform limited pulses of
approximately 40 fs duration. $f_o$ is measured via the typical $\it
f$-2$\it f$ interferometer technique and locked to a stable radio
frequency (rf) source which is fixed at a prescribed value or
scanned. $f_r$ is stabilized via a Cs-referenced low phase-noise rf
source and steered so as to maintain the appropriate absolute
frequency of the comb modes. In the following experiments, $f_r$ is
locked to 100.41356730 MHz, which for $f_o$ values of either 18.14
MHz or -31.86 MHz ensures two-photon resonance between the
$5S_{1/2}$ F=2 ground state and $5D_{5/2}$ F=4. For $f_o$ = 18.14
MHz, there is a comb mode resonant with the intermediate state
$5P_{3/2}$ F=3. In the other case of $f_o$ = -31.86 MHz, no mode is
resonant with $5P_{3/2}$ F=3, but rather all modes are symmetrically
detuned around this intermediate resonance. The signal is
proportional to the 5D population and is measured by photon counting
the 420 nm 5D-6P-5S cascade fluorescence. For appropriately chosen
values of $f_r$ and $f_o$, the signal after a sufficient number of
pulses to reach steady state excitation can be attributed to only
this $5D_{5/2}$ F=4 resonant hyperfine level \cite{Marian1}. After
many pulses all other 5D and 7S hyperfine states, although covered
by the 30 nm comb spectrum, are detuned many linewidths away from
the closest comb modes and as a result remain unexcited.

\begin{figure}[t]
\resizebox{7.0cm}{!}{
\includegraphics[angle=0]{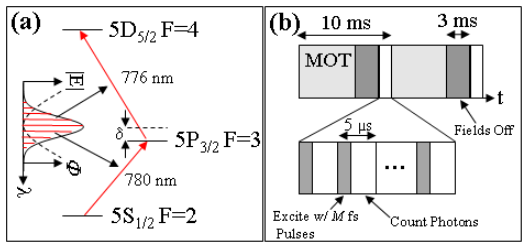}}
\caption{\label{Fig1}(color online) $\bf(a)$ Simplified energy level
diagram of the three relevant atomic states and frequency domain
representation of the chirped comb. $\delta$ is the detuning of the
intermediate state from half the two-photon transition frequency.
$\bf(b)$ Timing diagram used in the experiments.}
\end{figure}

The theory model describing the experimental results is based on
numerical solution of the Liouville equation for the density matrix
of the three level atomic system. Decoherence of the electronic
system is included via phenomenological decay terms due to the
natural linewidth of 6 MHz for the 5P and 660 kHz for the 5D atomic
states. The driving field is taken to be a linearly chirped pulse of
30 nm bandwidth at 778 nm and a transform limited peak field
strength of $E_o=10^7$ V/m.  The solution of the Liouville equation
is stepped through time via a fourth order Runge-Kutta algorithm.
After numerical solution for a single pulse, the density matrix is
advanced in time by approximately $1/f_r$ to the next pulse
analytically, then repeated for an arbitrary number of pulses. This
technique differs from some previous work on modeling two-photon
coherent control with femtosecond pulses as it includes effects of
atomic decoherence and the impulsive excitation approximation is not
used \cite{Felinto1,Girard1}. Our model includes only the $5S_{1/2}$
F=2, $5P_{3/2}$ F=3, $5D_{5/2}$ F=4 hyperfine levels. Because the
comb is tuned on resonance with this transition, the three-level
model is a good approximation to our experiment after several
pulses. However, for excitation by less than a few pulses the other
hyperfine states can not necessarily be considered off-resonance
because the frequency comb structure of the field is not manifest.

The data acquisition cycle (Fig. 1(b)) consists of loading the MOT,
shutting off all MOT-related fields, then exciting the free cold
atoms. A pockels cell with 8 ns rise time is used as a fast shutter
to transmit any desired number of pulses to the atoms. The
triggering for the pockels cell is directly from the laser
repetition frequency to ensure that the pockels cell is always fully
on or off when a femtosecond pulse is incident. Tunable linear
frequency chirp is generated by a double pass grating stretcher and
compressor which allows for chirp up to $\phi''$ = $\pm$ 2.5$\times
10^5 fs^2$, defined as $\phi''$ = $d\phi^2/d\omega^2$. Some higher
order dispersion is present, but our data is generally well modeled
by the inclusion of only quadratic spectral phase.  The MOT is
loaded for 6 ms, then the magnetic field is turned off for 3 ms
during which time an optical molasses remains, finally the atoms are
probed after all other fields are off. The timing diagram in (Fig.
1(b)) shows the atoms are excited by $\it M$ pulses, spontaneously
emitted photons are counted for 5 $\mu$s as the atoms relax, then
repeated 40 times per MOT.

We first demonstrate the coherent accumulation of a controlled
number of femtosecond pulses on the $5D_{5/2}$ F=4 population and
lineshape. Previous work on DFCS \cite{Marian1,Marian2} was
conducted only under steady state atomic excitation, i.e. excitation
times longer than the atomic coherence time. This is necessary for
spectroscopy with the highest possible resolution, but we
investigate properties of the transient femtosecond coherent
accumulation for pulse numbers, or equivalently excitation times
significantly less than the 240 ns $5D_{5/2}$ F=4 atomic lifetime.
The $f_r$ and $f_o$ of the comb are tuned to 100.41356730 MHz and
18.14 MHz respectively, which ensure that a comb mode is resonant
with the $5S_{1/2}$ F=2 to $5P_{3/2}$ F=3 one-photon transition and
another with the $5P_{3/2}$ F=3 to $5D_{5/2}$ F=4 transition (Fig. 2
inset). We split the probe laser and counter-propagate the two
intensity-matched beams through the atoms to balance the radiation
pressure and thus minimize any line broadening due to accumulated
Doppler shifts. For the chirped excitation experiments described
later the laser is not counter-propagated. The femtosecond probe
laser is focused with linear polarization to a diameter of 160
$\mu$m into the MOT, resulting in an electric field strength per
beam of approximately $10^7$ V/m. Figure 2 shows the total number of
photon counts (in squares) that occur after the $M^{th}$ pulse and
is a measure proportional to the total 5D population after
excitation by $\it M$ pulses. The dashed line in Fig. 2 is the
result from our theory model for low power multi-pulse excitation
such that there is no power broadening of the linewidth. In the
actual experiment the pulse intensity is high enough to excite a
significant 5D population and it saturates to its maximum value at a
smaller number of pulses than in the case of weak excitation (Fig.
2). This is the time domain picture of power broadening of the
transition linewidth due to the femtosecond probe laser. For pulse
numbers significantly smaller than the 5D coherence time the
population scaling can be fit well by a second order polynomial in
the pulse number or equivalently the accumulated area (inset Fig.
2). This coherent enhancement of the population versus pulse number
demonstrates the increased efficiency compared to excitation by a
single pulse of the same total power for the case of two-photon
absorption via a resonant intermediate state.

\begin{figure}[t]
\resizebox{6.5cm}{!}{
\includegraphics[angle=0]{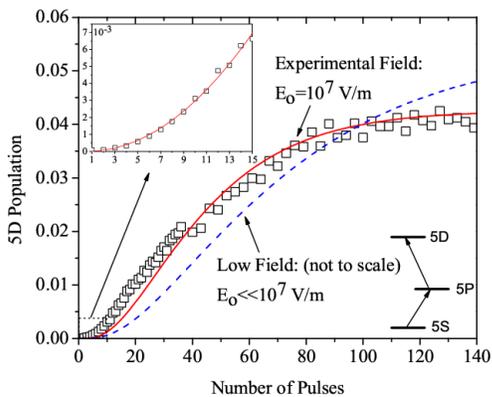}}
\caption{\label{Fig2}(color online) The on-resonance $5D_{5/2}$ F=4
fluorescence signal is measured (squares) at a field strength of
approximately $E_o=10^7$ V/m versus the number of applied pulses.
The solid (dashed) line is the theoretically predicted excited state
populations under experimental (asymptotically low) field strengths.
The inset shows the quadratic scaling for the first 15 pulses.}
\end{figure}

The coherent accumulation that enables the large enhancement of the
resonant 5D population simultaneously enables the high resolution
necessary for exciting a single 5D hyperfine level. In this part of
the experiment the lineshape is measured by scanning $f_o$ after
excitation from $\it M$ pulses to directly demonstrate the
resolution enhancement. As was mentioned previously the spectral
content of the field generated by a finite number of pulses is a
series of modes each with a linewidth inversely proportional to the
number of pulses coherently accumulated. Figure 3 illustrates this
effect through the lineshape and width of the measured $5D_{5/2}$
F=4 resonance after excitation by 10, 15, 20, and 80 pulses. The
asymptotic power broadened linewidth of 2 MHz occurs at 80 pulses
which is approximately the number of pulses the $5D_{5/2}$ F=4
population in Fig. 2 becomes saturated.

\begin{figure}[t]
\resizebox{6.7cm}{!}{
\includegraphics[angle=0]{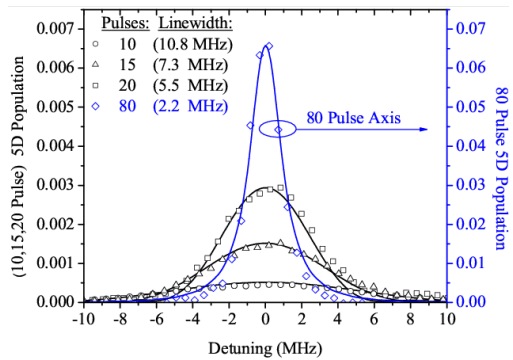}}
\caption{\label{Fig3}(color online) The $5D_{5/2}$ F=4 lineshape is
measured (symbols) versus the number of applied femtosecond pulses
(10, 15, 20, 80 pulses) with the corresponding FWHM linewidth in the
legend. The Lorenztian linewidth of 2.2 MHz after 80 pulses is the
asymptotic power broadened linewidth.}
\end{figure}

\begin{figure}[b]
\resizebox{6.7cm}{!}{
\includegraphics[angle=0]{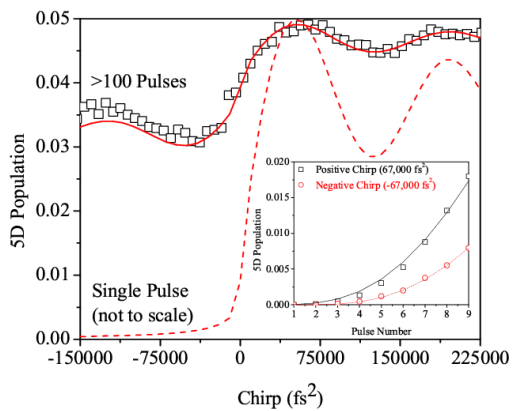}}
\caption{\label{Fig4}(color online) The $5D_{5/2}$ F=4 excited state
fluorescence is measured (squares) under steady state excitation
($>$100 pulses) versus the applied linear frequency chirp. The solid
line indicates the steady state theory and the dashed line is the
single pulse theory. Inset: The signal (symbols) versus pulse number
for two values of equal and opposite chirp, the solid and dashed
lines are theoretical predictions.}
\end{figure}

Some of the original work on coherent control has demonstrated the
modification of the quantum interference of the two-photon
transition in Rb and Cs by shaping a single amplified femtosecond
pulse \cite{Silberberg1,Noordam2,Girard1}. It was shown that for the
transition from 5S to 5D in Rb the excited state population was
essentially zero for large negative chirp, where higher frequencies
precede lower (see the single pulse theory in Fig. 4). For low field
strengths (in the absence of adiabatic following effects
\cite{Bergmann1,Noordam1}) and large negative chirps, this can be
understood conceptually as the front of the pulse is only resonant
with the 5P to 5D transition. Due to the fact that there is no
initial population in 5P a resonantly enhanced two-photon transition
can not occur, the only contribution can be from the frequencies
corresponding to approximately half the 5S-5D energy splitting.
Similar to previous experiments with single pulse excitation we
observe a characteristic oscillation of 5D population versus chirp
with a period given by $\phi''_{2\pi}$ = 2$\pi/\delta^2$ \cite
{Noordam2}; the definition of $\delta$ is given in Fig. 1(a). The
measured period of $\phi''_{2\pi}$ = 14.5$\times 10^4$ $fs^2$ agrees
well with the theory value of $\phi''_{2\pi}$ = 14.3$\times 10^4$
$fs^2$ for $5P_{3/2}$ F=3. However, for negative chirps
significantly different results are observed for multi-pulse
excitation as shown in Fig. 4. This difference is due to the 5S-5P
electronic coherence that persists for approximately 50 ns, or about
five inter-pulse periods. For negative-chirp pulses the intermediate
state can be excited by the low frequency end of the preceding pulse
and the high frequency beginning of the next pulse can complete the
two-photon transition. Figure 4 inset shows the 5D population at
$\phi''$ = $\pm$ 6.7$\times 10^4$ $fs^2$ versus the number of pulses
to illustrate that the scaling of 5D population is smaller for
negative chirp. We emphasize that unlike single pulse experiments
our results involve only three well resolved hyperfine levels chosen
such that hyperfine selection rules prohibit any transitions from
other states to $5D_{5/2}$ F=4. It is for this reason that we may be
attribute the measured oscillation period, $\phi''_{2\pi}$, to only
one intermediate state. This contrasts previous work where the
$5P_{1/2}$ states exhibit a shorter period, $\phi''_{2\pi}$ =
2.38$\times 10^3$ $fs^2$, oscillation superposed on the larger
$5P_{3/2}$ signal \cite{Girard1,Noordam1}.

\begin{figure}[t]
\resizebox{6.7cm}{!}{
\includegraphics[angle=0]{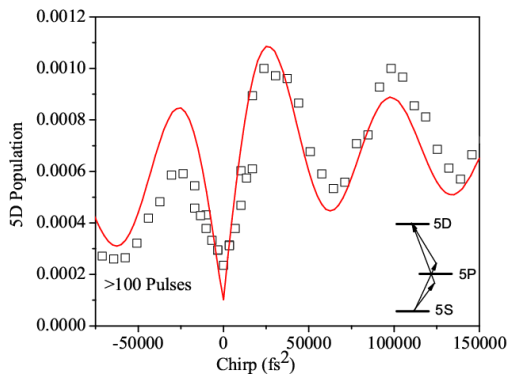}}
\caption{\label{Fig5}(color online) Measured fluorescence (squares)
and theoretical prediction (solid line) for the case of destructive
quantum interference versus linear frequency chirp. The inset shows
two of the many two-photon resonant comb mode pairs symmetrically
detuned from the intermediate $5P_{3/2}$ F=3 resonance, the closest
pair is $\pm$50 MHz ($\pm f_r$/2) detuned.}
\end{figure}

Our final experiment further demonstrates the versatility of the
high resolution frequency comb with spectral phase control by
significantly reducing the nonlinear absorption. The phase of the
two-photon transition amplitude for a pair of modes is a function of
the intermediate state detuning \cite{Ye3}. The detuning of the
closest mode to the intermediate $5P_{3/2}$ F=3 state can be tuned
from 0 MHz to at most $f_r$/2. In the case of all modes
symmetrically detuned around $5P_{3/2}$ F=3, the relative phase
between the $\pm$ 50 MHz detuned two-photon transition amplitudes is
$2 \times tan^{-1}$ (2$\times$(50 MHz/6 MHz)) $\simeq$ $173^0$,
between all other further detuned mode pairs it is $\simeq$ $180^0$.
This phase flip around the intermediate state results in nearly
complete destructive quantum interference of the two-photon
transition probability for transform limited pulses
\cite{Silberberg1}. Translating the entire comb by varying $f_o$ can
easily position all mode pairs symmetrically around the intermediate
state, while maintaining two photon resonance (Fig. 5, inset). As
shown in Fig. 5, this results in approximately 50 times less 5D
population than the previous case when a mode is tuned on $5P_{3/2}$
F=3 resonance. Linear frequency chirp introduces extra relative
phase between amplitudes detuned above and below the 5P resonance
and reduces the net destructive interference. As shown in Fig. 5,
the two-photon transition probability is modulated versus chirp,
with a minimum 5D population at zero chirp.

In conclusion, we have demonstrated transient coherent accumulation
via a controlled number of phase-coherent femtosecond pulses in an
atomic system. The application of spectral phase manipulation to
this multi-pulse regime enables fundamental coherent control
experiments to be performed at a resolution limited only by natural
resonance widths. The present work is a step towards combining pulse
shaping from the field of coherent control with femtosecond comb
technology. This research is expected to enable high resolution
coherent control, for example, in wavepacket shaping for molecular
photo-association \cite{kosloff1}.

We thank funding support from ONR, NSF, and NIST. We gratefully
thank A. Marian for the work on steady-state DFCS.

*Permanent address: Instituto de Fisica Gleb Wataghin, UNICAMP,
Campinas, SP, 13083-970, Brazil.

\end{document}